\documentclass{webofc}
\usepackage[varg]{txfonts}   % Web of Conferences font
\newcommand{\hGpc}{~h^{-1}~{\rm Gpc}}

\newcommand{\hMpc}{{\ifmmode{~h^{-1}{\rm Mpc}}\else{$h^{-1}$Mpc}\fi}}
\newcommand{\hkpc}{{\ifmmode{~h^{-1}{\rm kpc}}\else{$h^{-1}$kpc}\fi}}
\newcommand{\hMsun}{{\ifmmode{\,h^{-1}{\rm {M_{\odot}}}}\else{$h^{-1}{\rm{M_{\odot}}}$}\fi}}

\newcommand{\Mstar}{{\ifmmode{,M_{*}}\else{$M_{*}$}\fi}}
\newcommand{\Mhalo}{{\ifmmode{\,M_{\rm halo}}\else{$M_{\rm halo}$}\fi}}
\newcommand{\ltsima}{$\; \buildrel < \over \sim \;$}
\newcommand{\gtsima}{$\; \buildrel > \over \sim \;$}
\newcommand{\lsim}{\lower.5ex\hbox{\ltsima}}
\newcommand{\gsim}{\lower.5ex\hbox{\gtsima}}

\newcommand{\theth}{{THREEHUNDRED}\,}
\newcommand{\ahf}{\textsc{AHF}\,}

\newcommand{\gadgetx}{\textsc{Gadget-X}\,}
\newcommand{\simba}{\textsc{Gizmo-SIMBA}\,}
\newcommand{\gadgetmusic}{\textsc{Gadget-MUSIC}\,}

\begin{document}
\title{The \theth project: the effect of baryon processes at galaxy cluster scale}
%
% subtitle is optionnal
%
%%%\subtitle{Do you have a subtitle?\\ If so, write it here}

\author{\firstname{Weiguang} \lastname{Cui}\inst{1}\fnsep\thanks{\email{weiguang.cui@ed.ac.uk}} 
%     \and 
% 	\firstname{Second author} \lastname{Second author}\inst{2} 
% 	\and
% 	\firstname{Third author} \lastname{Third author}\inst{3}
        % etc.
}

\institute{Institute for Astronomy, University of Edinburgh, Royal Observatory, Edinburgh EH9 3HJ, UK
% \and
%           the second here 
% \and
%           Last address
          }

\abstract{%
The role of baryon models played in hydrodynamic simulations is still unclear. Future surveys that use cluster statistics to precisely constrain cosmology models require a better understanding of that. With the hydro-simulated galaxy clusters from different baryon models (\gadgetmusic, \gadgetx and \simba) from the \theth project, we can look into more details of this question. We find that the galaxy cluster mass change due to different baryon models is at few percent level. However, the mass changes can be positive or negative, which is depending on the baryon models. Such a small mass change leaves a weak influence (slightly larger compared to the mass changes) on both the cumulative halo numbers and the differential halo mass function (HMF) above the mass completeness. Agreed to the halo mass change, the halo mass (or HMF) can be increased or decreased with respect to the dark-matter-only (DMO) run depending on the baryon models.
}
\maketitle
\section{Introduction}
\label{intro}
More than a decade before, theoretical studies on structure and galaxy formations are relying N-body DMO simulations, see the well-known Millennium simulations \citep{Springel2005N} and the MultiDark simulations \citep{Klypin2016}, for example. As we can only directly observe the galaxies in the sky and simulations only contains gravitational bound objects -- dark matter halos, there is a missing connection between observation -- galaxies, and theoretical prediction -- dark matter halos. Thus, numerous techniques, such as hydrodynamic simulations, semi-analytic models, empirical forward modelling, subhalo abundance modelling and halo occupation models (from more physical models to more empirical models), are developed to bridging this connection \citep[see][for a recent review]{Wechsler2018}. Ideally, we would like to model the baryon process as physical as we could. However, limited to the current computation power, unknown baryon processes or uncompleted baryon models implemented in the simulation codes \citep[see][for the current development of baryon models in hydrodynamic simulations]{Somerville2015} as well as the dependence on the simulation resolution\citep[][]{Ludlow2019}, we are still facing difficulties to fully model galaxy formation at the non-linear scales. Nevertheless, the hydrodynamic simulation with baryon evolved simultaneously inside is the only way to understand the difference between these DMO simulations and the effects of baryons on them. Numerous work has studied the effects from different aspects \citep[see][for a review]{Cui2017book}. However, no clear agreement has been reached, especially at the non-linear scale. 

Galaxy clusters as the one of the independent constraints on cosmology model parameters, will be able to provide precise values of $\sigma_8$ and $\Omega_m$ with errors at around 0.1 percent from the next-generation space telescopes \citep[see][for example]{Sartoris2016}, such as EUCLID, CSST. Therefore, theoretically understanding the cluster masses, especially their changes according to different baryon models, is essential to accurately constrain cosmology parameters \citep[see][for a detailed discussion]{Debackere2021}. In this short article, we will probe this question, focusing on the galaxy cluster scale with clusters simulated with different codes from the \theth project\citep{300Cui2018}. The outline of this paper is following: the \theth project is introduced at Sect.~\ref{sec:300}; the effects of baryons on the halo mass and HMF are presented in Sect.~\ref{sec:hm}; we finally list the conclusions and discuss our results in Sect.~\ref{sec:conc}.

\section{The simulated galaxy clusters from the \theth project} \label{sec:300}

The \theth project \citep{300Cui2018} is a re-simulation of 324 most massive galaxy clusters ($M_{vir} > 8\times 10^{14}\hMsun$)\footnote{The halo mass is defined as the mass enclosed inside an overdensity of $\delta$ times the critical density of the universe: $\delta = \sim 98$ for virial mass and $M_{200, 500}$ is with $\delta = 200, 500$ respectively.} from the MultiDark simulation (MDPL2, \cite{Klypin2016}, also refers as the DMO run in this paper) which utilises the cosmological parameters form the Planck mission \citep{planck2016}, and has a periodic cube of comoving length $1 \hGpc$ containing $3840^3$ DM particles with a mass of $1.5 \times 10^9 \hMsun$. Each cluster lies in the highest resolution region of a comoving radius of $15 \hMpc$ (over 5 times $R_{200}$) for re-simulations with different baryonic models: \gadgetmusic\citep{MUSICI}, \gadgetx\citep{Rasia2015,Beck2016}, \simba(\cite{Dave2019} and Cui et al. 2021 in prep. for more details). 

These re-simulation regions are generated with the parallel {\textsc GINNUNGAGAP} code\footnote{\url{https://github.com/ginnungagapgroup/ginnungagap}}: the highest resolution Lagrangian regions share the same mass resolution as the original MDPL2 simulation with gas particles ($M_{gas} = 2.36\times10^8 \hMsun$) split from DM particles. The outside regions are degraded in multiple layers (with a shell thickness of $\sim 4 \hMpc$) with lower mass resolution particles (mass increased by eight times for each layer) that eventually provide the same tidal fields yet at a much lower computational costs. 
The halo catalogue used in this paper is generated with \ahf \citep{Knollmann2009} and we mainly focusing on halo masses within two interesting overdensities: 200 and 500, at $z=0$, in this paper. 

Benefited from the large volume of these re-simulated cluster regions, detailed relations between the central cluster and the filaments connecting to it have been studied \citep{300Rost2021,300Kuchner2020,300Kuchner2021}. Furthermore, the cluster backsplash galaxies \citep{300Haggar2020,300Knebe2020} and shock radius \citep{300Baxter2021} have also been well addressed. The advanced baryon models in hydrodynamic simulations allow us to perform a detailed investigation on the cluster properties, such as profiles \citep{300Mostoghiu2019,300Li2020}, substructure and its baryonic content \citep{300Arthur2019,300Haggar2021,300Mostoghiu2021,300Mostoghiu2021b}, dynamical state and morphology \citep{300Capalbo2021,300DeLuca2021}, cluster (non-)thermalization \citep{300Sayers2021,300Sereno2021}, the fundamental plane \citep{300Diaz-Garcia2021}, and the cluster mass biases \citep[][for hydrostatic-equilibrium assumption]{300Ansarifard2020}, \citep[][for sigma-mass relation]{300Anbajagane2021} and \citep[][for weak-lensing]{300Herbonnet2021}. Lastly, comparing to the void/field region runs in this project allows us to study the effect of environment \citep{300Wang2018}; including the self-interacting dark matter run allow us to constrain the dark matter cross-section\citep{300Vega-Ferrero2021}; It further help us to examine the chameleon gravity\citep{300Tamosiunas2021}.

\section{The effects on halo mass and HMF} \label{sec:hm}
% For figure with sidecaption legend use syntax of figure
% \begin{figure}
% % Use the relevant command for your figure-insertion program
% % to insert the figure file.
% \centering
% \sidecaption
% %\includegraphics[width=5cm,clip]{tiger}
% \caption{Please write your figure caption here}
% \label{fig-3}       % Give a unique label
% \end{figure}
\begin{figure}
\centering
% Use the relevant command for your figure-insertion program
% to insert the figure file. See example above.
% If not, use
\sidecaption
\includegraphics[width=8cm,clip]{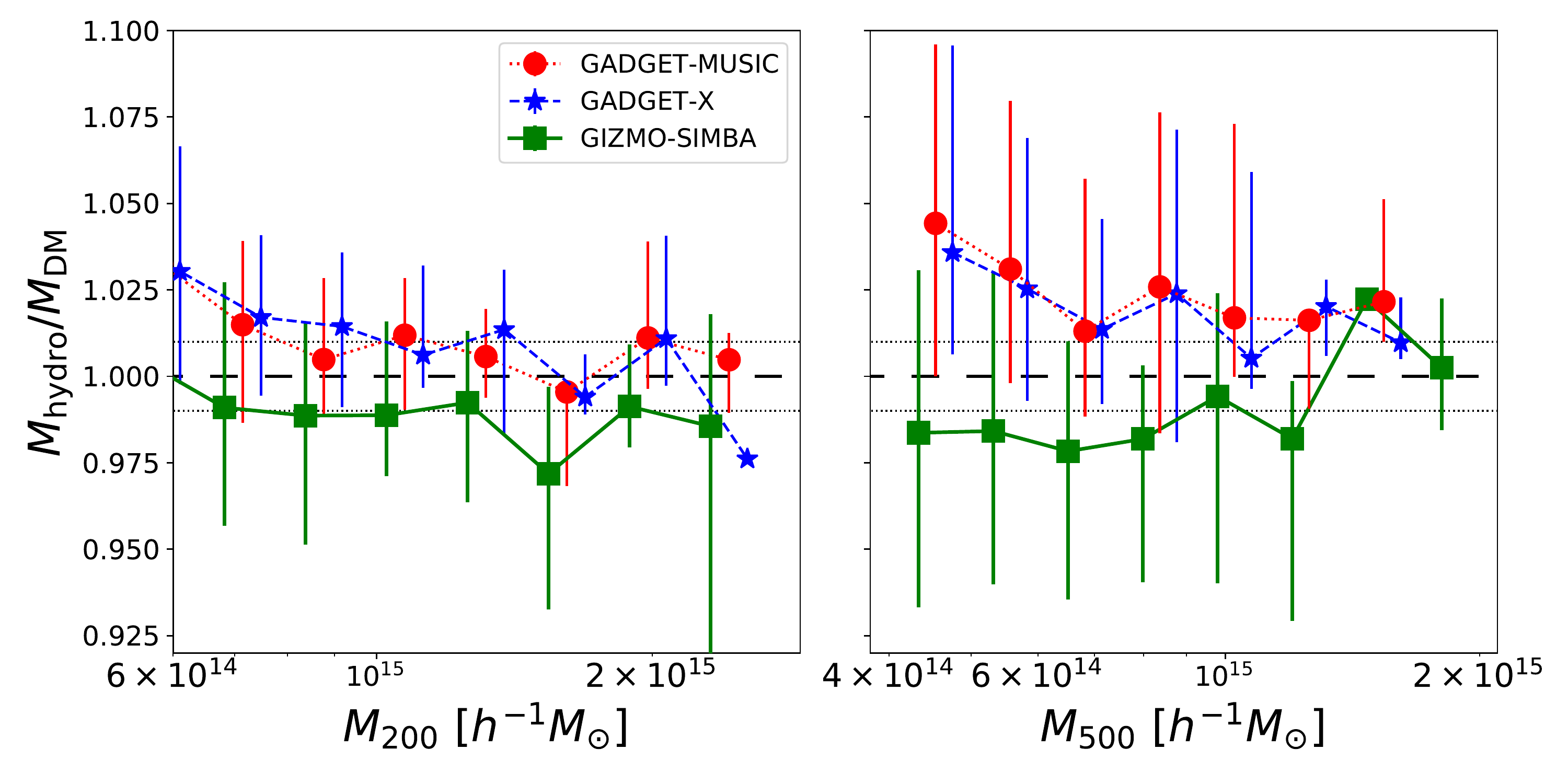}
% \vspace*{5cm}       % Give the correct figure height in cm
\caption{The halo mass differences at $M_{200}$ (left panel) and $M_{500}$ (right panel)  with respect to the DMO run. X-axis shows the halo masses from the DMO run. As indicated in the legend, different symbols with colours show different simulation runs. The errorbars show the $16^{th} - 84^{th}$ percentiles. Only the mass-complete sample is presented here. The horizontal dashed line marks the unit value of 1 with the two dotted lines show the 1 percent difference in mass.}
\label{fig-1}       % Give a unique label
\end{figure}

\begin{figure*}
\centering
% Use the relevant command for your figure-insertion program
% to insert the figure file. See example above.
% If not, use
\sidecaption
\includegraphics[width=8cm,clip]{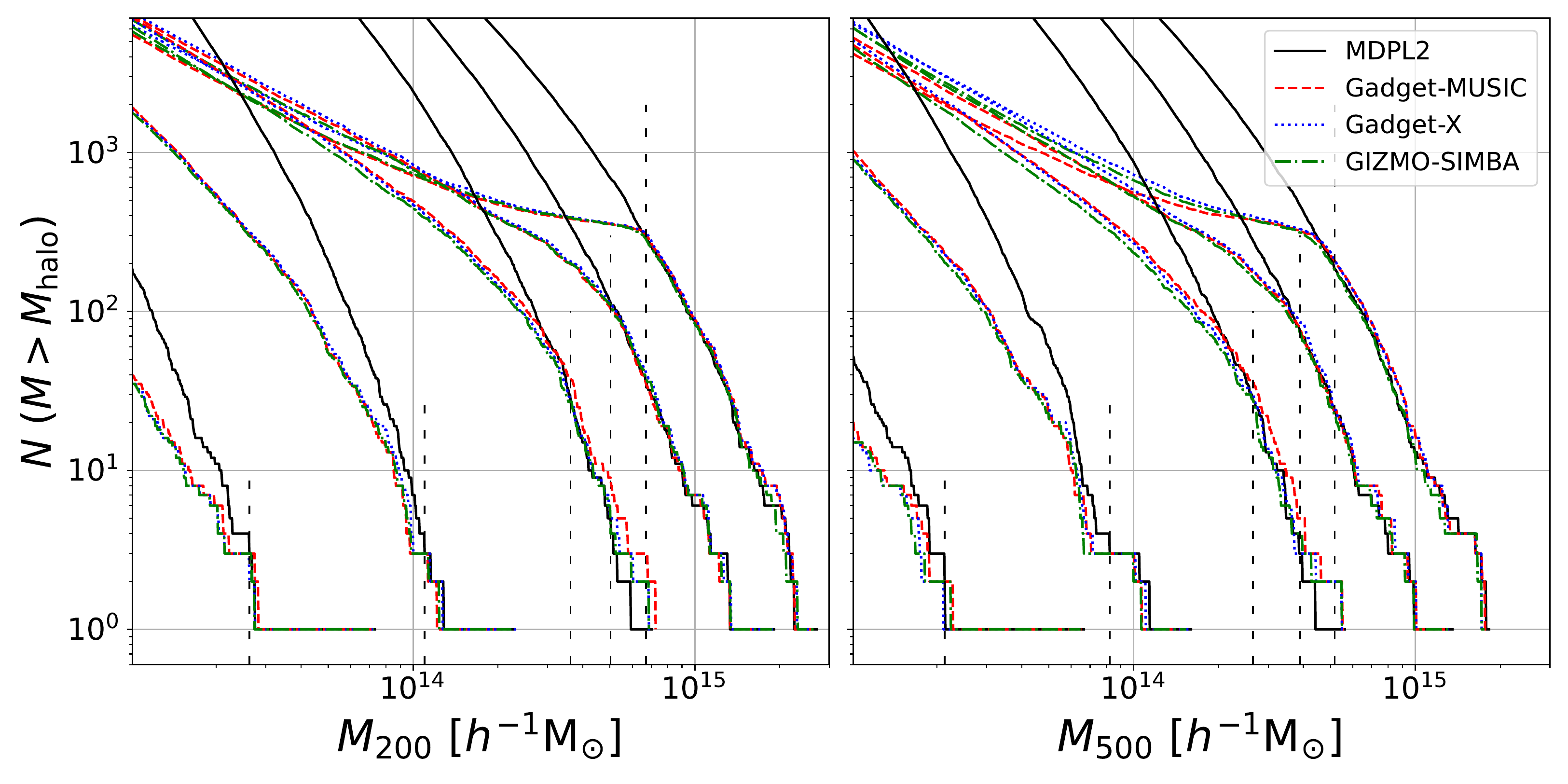}
% \vspace*{5cm}       % Give the correct figure height in cm
\caption{The cumulative HMF from MDPL2 (solid black lines), \gadgetmusic(red dashed lines), \gadgetx(blue dotted lines) and \simba(green dot-dashed lines). Left panel shows the results of $M_{200}$ with $M_{500}$ is presented on the right panel. In each panel, the results from $z=4.0, 2.3, 1.0, 0.5$, and 0.0 are presented from left to right, respectively. Vertical dashed lines show the mass-completeness limitation at each redshift.}
\label{fig-2}       % Give a unique label
\end{figure*}

% For tables use syntax in table~\ref{tab-1}.
\begin{table}
\centering
\caption{The mass-complete samples at different redshifts. The second row shows the mass completeness thresholds for $M_{200}$ with the numbers of clusters above that threshold in MDPL2, \gadgetmusic, \gadgetx and \simba, respectively. Similar results at $M_{500}$ are at bottom.}
\label{tab-1}       % Give a unique label
% For LaTeX tables you can use
\begin{tabular}{llllll}
\hline
Redshift & z=0.0 & z=0.5 & z=1.0 & z=2.3 & z=4.0 \\\hline
$M_{200} ~ [10^{14} \hMsun]$ & 6.71 & 5.02 & 3.62 & 1.10 & 0.26 \\
Number of clusters & 284/300/308/282 & 112/104/110/108 & 26/38/27/27 & 3/3/3/3 & 3/3/2/3 \\\hline
$M_{500} ~ [10^{14} \hMsun]$ & 5.16 & 3.90 & 2.65 & 0.82 & 0.21 \\
Number of clusters & 184/208/206/179 & 73/76/85/73 & 28/36/30/28 & 3/3/3/3 & 1/2/1/2 \\\hline
\end{tabular}
% Or use
% \vspace*{5cm}  % with the correct table height
\end{table}
\begin{figure}
\sidecaption
\includegraphics[width=8cm,clip]{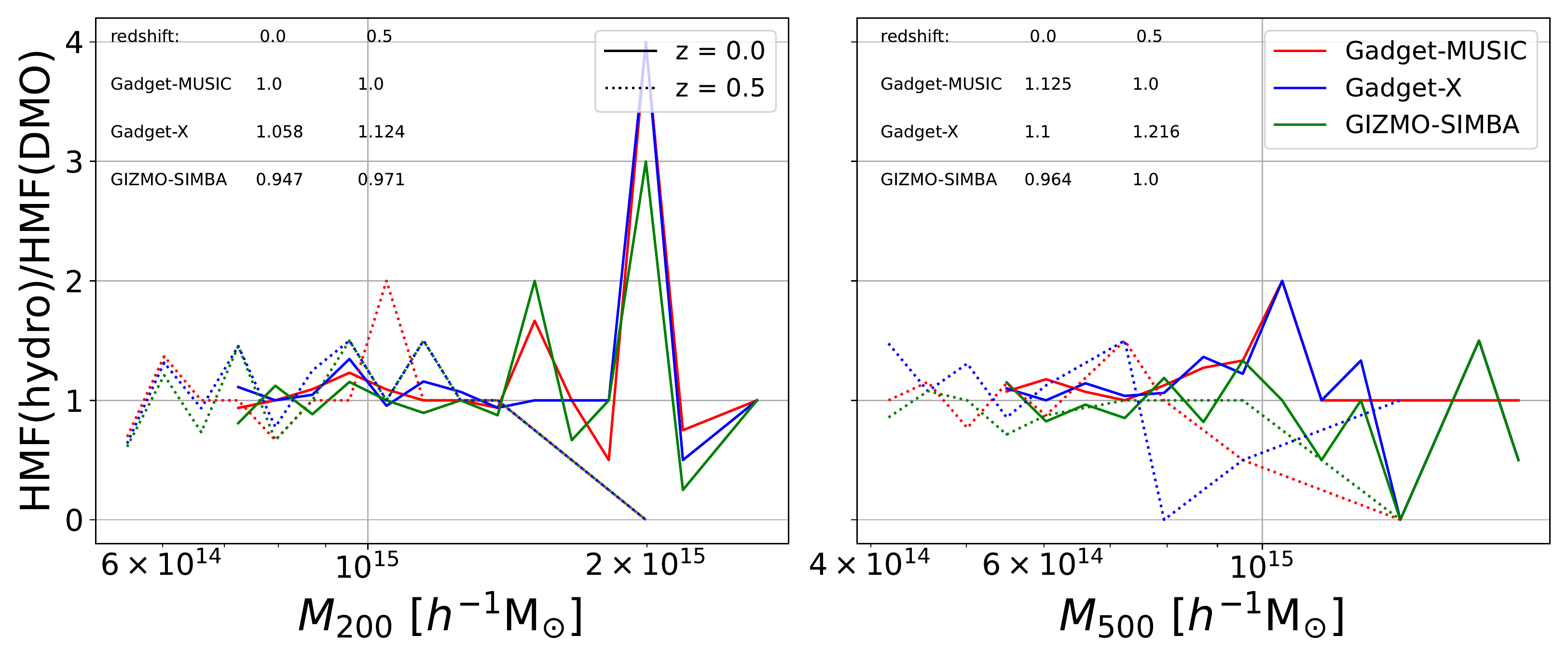}
% \vspace*{5cm}       % Give the correct figure height in cm
\caption{The differential HMF differences with respected to the one from MDPL2 above the mass completeness. The same mass bin with size of 0.05 in logarithm space is applied. Left panel shows the results of $M_{200}$ with $M_{500}$ is presented on the right panel. In each panel, only the results from $z= 0.5$ (dotted lines), and 0.0 (solid lines) are presented due to the statistic limits. The median changes for all bins with $\geq 8$ clusters are shown in the top left corner of each panel.}
\label{fig-3}       % Give a unique label
\end{figure}

The effect of baryons on HMF is simply a consequence of the halo mass change. Therefore, we need to firstly investigate the halo mass changes due to the effects of baryon models. This requires a match of the galaxy cluster between different runs. For each DMO cluster, we simply use the cluster centre positions ($\Delta_r < 200 \hkpc$) and their masses ($\Delta_M/M_{200} < 0.1$) to find its corresponding ones. 

Figure~\ref{fig-1} shows the relative mass differences between the hydro-simulated clusters and their DMO counterparts. As there are many clusters, we simply bin them in the DMO cluster mass and show the median values with errorbars. At $M_{200}$, both \gadgetmusic and \gadgetx tend to have slightly (at about 1 percent) higher mass than \simba which seems about 1 percent lower than the DMO halo mass. \gadgetx shows weak dependence on halo mass, while \simba does not within this mass-complete range. At $M_{500}$, all the three hydro runs tends to deviate from the DMO halo mass larger towards opposite directions, besides the highest halo mass end. It also worth to note that the errorbars are slightly larger compared to the ones in the left panel. 

At lower halo mass $M_{halo} \lesssim 10^{14} \hMsun$, the hydro-simulated halo tends to gain mass without AGN feedback \citep[see][for example]{Cui2012b}. However, it is well-known that it is very hard to cease the star formation in galaxies without AGN feedback which will result in unrealistic galaxy properties compared to observation. Including AGN feedback, the hydro-simulated halo tends to lose mass compared to the DMO run \citep[see][for example]{Cui2014,Castro2021}. At more massive end, this requires a very large volume simulation or zoomed-in simulations like the \theth to provide statistical information. It seems that the baryon models leave a weak influence ($\sim$ a few percents) on changing these massive cluster masses.

We continue to show the cumulative HMFs in figure~\ref{fig-2} for these different runs. The mass completeness of our cluster sample is shown by vertical dashed lines, which is determined by the crossing point between the \simba and MDPL2 lines. More details about the mass-complete sample can be found in table~\ref{tab-1}. Note that these values are slightly different to \cite{300Cui2018} which used the crossing point between \gadgetx and MDPL2. It is clear that the difference between these cumulative HMFs is very small for the mass-complete sample.

Lastly, in figure~\ref{fig-3}, the relative differences of these differential HMFs with respective to the DMO run seem to wiggle around 1 with no clear discrepancy within this limited mass range. Calculating the median raitos for all the mass bins with more than 8 clusters, we notice a few to about 10 percent changes depending on the simulations with lower values from \simba and higher values for \gadgetx. Similar to the halo mass changes, different baryon models can also change the HMF towards the same opposite directions.

\section{Conclusion and discussion} \label{sec:conc}

Use the hydro-simulated galaxy clusters from the \theth project, which was performed using different simulation codes, we investigate the effects of baryon models on the halo mass and HMF by comparing to their DMO counterparts. Limited to our mass-complete sample, we can only statistically present the results at the massive cluster mass scale. We find that the cluster mass changes due to different baryon models is at few percent level. There is a slightly larger deviation for $\Delta_{M_{500}}$ than $\Delta_{M_{200}}$. The cluster mass can be increased or decreased with respect to the DMO run and that is model dependent. Such a small change in mass leaves weak influence on the HMF (although slightly higher fractions compared to mass changes). However, different models can alter the changes into different directions of increase or decease, the same as the mass changes.

The effect of baryons on halo mass is tightly connect with the halo mass definition.
In this paper, we adopt the commonly used halo masses $M_{200}$ and $M_{500}$ which are based on the critical density of the universe. Within a fixed radius (not very large), we can expect that baryons condensed in the centre of the cluster will result in a high concentration, thus an increase in mass. While this may result a steep slope and the 200, 500 $\times \rho_{crit}$ can be reached at a shorter distance. Hence, maybe a lower $M_{200, 500}$. Therefore, it is also not surprising to see that there is less mass change for $M_{200}$ than $M_{500}$ as shown in figure~\ref{fig-1} \citep[see also][]{Cui2012b,Cui2014}. We note here that the FoF (Friend-of-Friend) halo definition seems to have a similar problem \citep[see][for example]{Cui2014}. Furthermore, the effect of baryons may change the centre of the halo \citep{Cui2016b}. Although this may not happen very frequently, the change due to this mis-centring problem is generally believed to bias towards a lower halo mass.

\section{Acknowledgement}
The author would like to thank the amazing organisers of the NIKA2 conference and the \theth project for contribution and collaboration. This work can not be finished without them. WC is supported by the STFC AGP Grant ST/V000594/1 and the science research grants from the China Manned Space Project with NO. CMS-CSST-2021-A01 and CMS-CSST-2021-B01.

\bibliography{sample,paper}
%
% Non-BibTeX users please use
%
% \begin{thebibliography}{}
% %
% % and use \bibitem to create references.
% %
% \bibitem{RefJ}
% % Format for Journal Reference
% Journal Author, Journal \textbf{Volume}, page numbers (year)
% % Format for books
% \bibitem{RefB}
% Book Author, \textit{Book title} (Publisher, place, year) page numbers
% % etc
% \end{thebibliography}

\end{document}